\documentclass[lnicst]{svmultln}
\usepackage{graphicx}
\usepackage{booktabs}
\usepackage{xspace}
\usepackage{multirow}
\newcommand{\Eg}{E.g.,\xspace}
\newcommand{\eg}{e.g.,\xspace}

\usepackage{makeidx}  
\begin{document}
\mainmatter              
\title{SDN-Assisted Network-Based Mitigation of Slow DDoS Attacks}
\titlerunning{SDN-Assisted Network-Based Mitigation of Slow DDoS Attacks}  
%
\author{Thomas Lukaseder \and Lisa Maile \and Benjamin Erb \and Frank Kargl}
\authorrunning{Thomas Lukaseder et al.}   
%
%
\institute{Institute of Distributed Systems, Ulm University, Albert-Einstein-Allee 11, 89081 Ulm\\
\email{<firstname>.<lastname>@uni-ulm.de},\\
\texttt{http://uni-ulm.de/in/vs}}

\maketitle              

\begin{abstract}        
Slow-running attacks against network applications are often not easy to detect, as the attackers behave according to the specification.
The servers of many network applications are not prepared for such attacks, either due to missing countermeasures or because their default configurations ignores such attacks.
%
%
The pressure to secure network services against such attacks is shifting more and more from the service operators to the network operators of the servers under attack.
Recent technologies such as software-defined networking offer the flexibility and extensibility to analyze and influence network flows without the assistance of the target operator.
%
%
%
%

%
Based on our previous work on a network-based mitigation, we have extended a framework to detect and mitigate slow-running DDoS attacks within the network infrastructure, but without requiring access to servers under attack.
We developed and evaluated several identification schemes to identify attackers in the network solely based on network traffic information.
We showed that by measuring the packet rate and the uniformity of the packet distances, a reliable identificator can be built, given a training period of the deployment network.


\end{abstract}


\section{Introduction}

On the Internet, Distributed Denial of Service (DDoS) attacks represent common attack mechanisms that aim for the availability of their target. For this purpose, DDoS attackers drain relevant resources of the victims (\eg network bandwidth or computing power of a host), eventually rendering their services unavailable for the users. A key to these attacks is the imbalance between resources invested by the attackers and resourced drained at the targets at the expense of the victim.
Some DDoS mechanisms exploit properties of network protocols (\eg SYN flooding during TCP connection establishment or amplification attacks with DNS replies), while others rely on excessive traffic generated by a large number of attacking entities (\eg by leveraging botnets). Most often, a combination of both is used.

Slow-running DDoS attacks represent a specific attack mechanism that takes advantage of the properties of many connection-oriented, request/response-based client/server protocols, especially HTTP. 
Instead of simply flooding the victim, a malicious client of a slow-running DDoS attack spawns connections to the target server and tries to keep these connections established as long as possible. At the same time, the client tries to minimize its own network bandwidth for those connections. Therefore, the attacker delays the completion of its initial request by purposefully fragmenting and trickling a valid request with extended periods of no transmissions. Such slow connections then tie up finite resources at the victim's server (\eg threads for request handling) and eventually hinder the service availability for legitimate users.

So far, countermeasures against slow-running DDoS attacks have been primarily considered for the actual servers under attack, as we highlight in Section~\ref{sec:background}.
We argue that this host-based approach is not sufficient in several scenarios and opt for a network-based approach instead.
In particular, a host-based approach requires appropriate adaptations to the individual server setups and configurations by all of their operators. 
In a network-based approach, the network of the target actively provides detection and mitigation capabilities, independent of the actual victim server.  
The main contribution of our work is presented in Section~\ref{sec:framework}, when we adapt our network-based, SDN-assisted DDoS mitigation framework to mitigate slow-running DDoS attacks. Our framework enables us to protect arbitrary servers in our network against such attacks, without any changes to the servers.
In Section~\ref{sec:evaluation}, we evaluate the capabilities of our framework using real-world network traffic.
Next, we give an outlook and point to practical implications when applying network-based mitigation schemes against slow-running DDoS attacks and summarize our findings in Section~\ref{sec:conclusion}.


\section{Background}
\label{sec:background}

Unlike many other DDoS attack variants, slow-running attacks do not rely on excessive network traffic to bring down a target host. Instead, a number of tentatively valid connections with very low bandwidth are used to drain server resources over time. For the detection and mitigation of such connections, suspicious connection properties must be identified. 
Because it is very difficult to distinguish attackers from regular slow clients, the detection of slow DDoS attacks is often challenging.

\subsection{Slow DDoS Attacks}

Slow DDoS attacks target application-level protocols that rely on a connection-oriented transport protocol (\eg TCP) and use client/server-based architectures.
As most of these protocols employ a request/response-based message exchange pattern, we do only consider this pattern. 

Under the premise that a server handles an incoming connection by providing a dedicated resource (\eg a request handler thread or process), it is the aim of a slow DDoS attack to deplete the pool of available connection resources. Note that these resources still remain idle most of the time, as the mechanism is attacking their availability, not their utilization.

\subsubsection{General Mechanism}
Given a generic connection-oriented, request/response-based application protocol, a client first establishes a connection, then sends (1) a request with its headers, optionally followed by (2) a request payload. After fully receiving the request, the server handles the request and replies with a response, consisting of (3) response headers and potentially (4) a response payload.

In either four of these steps, an attacker can deliberately slow down communication. While sending the request, the attacker can use very low transmission bandwidths and introduce artificial delays between sending chunks of data on the application level. When receiving a response, the attacker can delay and reduce the speed of the read operations, effectively announcing small receiving buffers and hence forcing the server to also use low bandwidths.

Apart from this agnostic approach, attackers can also exploit explicit properties of the application-level protocol to delay transmissions, prevent timeouts, or render a detection difficult due to valid behavior in line with the protocol specifications.

\subsubsection{Slow DDoS Attacks against HTTP}
Although slow-running DDoS attacks work against other protocols such as IMAP, SMTP, or FTP, the HTTP protocol is the most prominent victim of this attacking scheme yet. 

\paragraph*{Slow Header HTTP Attack} This attack is also known as slowloris~\cite{Slowloris} and is the predominant slow HTTP attack. It was successfully used in 2009 against Iranian government servers~\cite{Zdrnja2009}.
In the slow header HTTP attack, a malicious client starts with a regular HTTP request line. After that, the client waits a certain period of time before it sends an additional custom request header (\eg \texttt{"X-abcd: 1234}"). 
The client then waits another period and repeats the previous step with another random custom header.
According to the specification of HTTP~\cite{rfc2616}, clients are allowed to add such custom headers. 
This mechanism does not only slow down the initial request, in fact, it does not terminate the request at all. Unless the server applies countermeasures such as maximum request duration time, an ongoing slow request can bind server resources for an arbitrary period.

\paragraph*{Slow Body HTTP Attack} This variant is also known as the slow POST attack, as it relies on the HTTP \texttt{POST} method. This method allows the client to submit a request entity such as form data or file to be uploaded. While regular behavior is used for the request header, the attacker either slows down the transmission of the request entity, or provides a \texttt{Content-Length} which is deliberately larger than the actual entity. In turn, this requires the server to wait for additional data. Alternatively, an attacker can use the chunked transfer encoding mode in order to send arbitrarily slow chunks of a request entity.
 
\paragraph*{Slow Read HTTP Attack} In this variant, the attacker requests a large resource using a regular HTTP request~\cite{Park2014}. Once the server starts to send the HTTP response entity, the attacker consumes the incoming stream at an extremely slow rate, which forces the HTTP server to slow down the transmission due to the small receive buffer resp. full window~\cite{Tayama2018}. This attack requires much more resources from the attackers as the packets from the server need to be acknowledged and is therefore less common than the aforementioned attacks.

\subsection{Countermeasures against Slow DDoS Attacks}

There are commercial vendors claiming that their tools are able to mitigate slow-running attacks. Unfortunately, these vendors do not publish their mechanisms, nor are their tools freely available for open, comparative analyses. However, there are scientific publications in the area of our research that we elaborate on in the following.

While it is rather easy to detect slow-running connections, it is very difficult to determine whether these connections are from valid clients with bad network connections, or from malicious clients that execute an attack.
Countermeasures conducted directly by the server under attack have been receiving the majority of attention in the literature (\eg \cite{Hirakawa2016,evalControls,Tripathi2016}). As server applications terminate connections on the application layer, host-based mechanisms can take advantage of protocol-specific properties and metrics to estimate malicious behavior. \Eg a webserver can specify limits for the minimum data rate required for a client when sending an HTTP request. It can also use more aggressive timeout values for the initial HTTP request lines, subsequent header lines or chunks of HTTP messages. A drawback of host-based mechanisms is the fact that attack victims and attack deflectors are the same instances.

We opt for a network-based approach which provides protection for threatened services on a network level. 
So far, only Hong et al.~\cite{Hong2017} have suggested a network-based defense method against Slow HTTP DDoS attacks by using SDNs. Their method introduces an SDN-based defense application that is triggered by a webserver, but then handles potentially malicious HTTP traffic instead of the webserver.
The approach relies on assistance by the webserver under attack, as the webserver actively initiates the attack check routine and forwards message fragments to the defense application and requires access to the application level payload.

Unlike the method of Hong et al., our approach does not require any active assistance of the threatened servers. Instead, we only probe the servers in a way which is transparent to the servers and does not require any modifications to the services.

\subsection{Network-Based Mitigation}
\label{sec:networkbased}

Mitigation of DDoS attacks can be achieved at different locations in the network. On the one hand, the target host itself can deploy mitigation mechanisms. As previously mentioned, there are host-based mitigation mechanisms against slow attacks. However, these mechanisms require the administrator of the target service to become active. Many service operators do not have the resources or knowledge to mitigate attacks on their end. Therefore, mitigating of DDoS attacks is often handled by network operators and offered as a service in form of DDoS Protection Services that become increasingly popular~\cite{Jonker2016}.

A differentiation amid network-based mitigation mechanisms can be made whether the mechanism is transparent for the target host and acts autonomously or whether the target has to actively request the mitigation from the mitigation service providers and has to cooperate for the mitigation to be effective.
The aforementioned mechanism by Hong et al. is a network-based mitigation mechanism based on SDN against slow HTTP attacks that identifies attackers based on timeouts. The proposed mechanism relies on the cooperation of the target host~\cite{Hong2017}.

Fayaz et al.~\cite{bohatai} propose Bohatai as an ``Flexible and elastic DDoS defense'' tool for SDN, NFV, network-based mitigation system. However, they do not consider slow attacks.

In prior work, we proposed a network-based mitigation system utilizing SDN. The system is transparent to the target system and acts autonomously. However, it is only capable of identifying attackers in flooding attacks~\cite{LCN17}. The attack detection and mitigation mechanism is attack agnostic and can therefore also be used for slow attacks. 


\section{Framework}
\label{sec:framework}

In the following, we describe the attacker model underlying our system. Based on that, we describe the architecture of the system on a conceptual level followed by a description of the schemes to identify attackers and the description of our implementation.

\subsection{Attacker Model}

The attackers in these scenarios have access to a large number of distributed network resources (e.g. a bot net). Attackers making use of DDoS attacks are differing greatly in resources and technical knowledge. We differentiate between two types of attackers.
On the one hand, the simple attacker. The simple attacker uses tools that are readily available for their attacks and several of those exist for slow attacks. We also make use of these very tools and the options they provide to emulate the behavior of this attack type.
The second attacker model features a more sophisticated attacker that knows how the attacks work, understands how these attacks are usually mitigated, and built their own or made adjustments to the attacking tool to circumvent detection.


\subsection{Architecture}

\begin{figure}
    \centering
    \includegraphics[width=.8\textwidth]{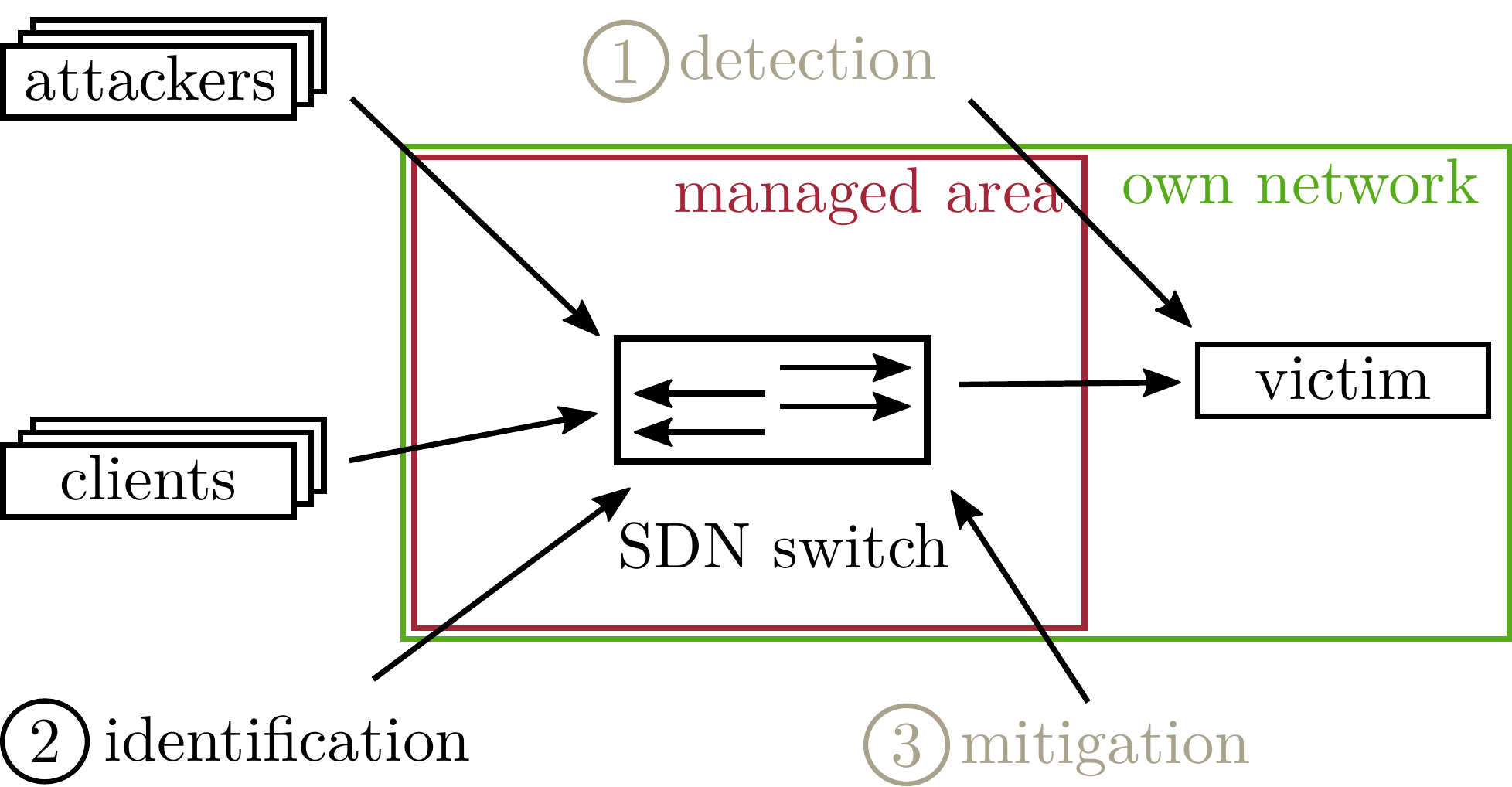}
    \caption{Sketch of the framework architecture, as described in full detail in \cite{LCN17}. The contribution of this work focuses primarily on phase 2, the identification of attackers.}
    \label{fig:arch}
\end{figure}

The architecture of the mitigation system is based on our previous work. In~\cite{LCN17}, we have introduced our framework and the underlying concepts. We have also implemented the framework, including identification schemes for flooding attacks. In an evaluation, we have shown the system's capabilities to mitigate TLS, HTTP, and SYN flooding attacks measuring detection times, times until the last attacker was blocked, and the overall downtime of the server. The system has proven to be capable of detecting and mitigating these attacks at 10~Gbps throughput on commodity hardware with few falsely identified clients. The concept is shown in Figure~\ref{fig:arch}.

We have divided the functionality of our framework in three phases. The detection phase (1) responsible for detecting an attack, the identification phase (2) identifying the attackers, and the mitigation phase (3) where the attackers are excluded from the network. The system is a network-based mitigation system capable of detecting arbitrary DDoS attacks and mitigating flooding attacks.

In the \emph{detection phase} (1), the system observes the network to find attackers. For that, the status of servers that are under protection of the system is monitored. If a server cannot be reached, the system assumes an attack and starts to analyze the traffic data in the identification phase.

The \emph{identification phase} (2) for the target system is activated to identify the attackers. For this, the SDN controller instructs the SDN-enabled switches upstream of the target to mirror the target's incoming traffic and to forward it to a network monitor. The monitor then identifies possible attackers by applying the different identification schemes and relays this information to the SDN controller. For flooding attacks, we use a scoring-based identification scheme. Clients are given a score dependent on the assumed load a packet of that type would inflict on the target system. For example, an \emph{IP} packet would give one point, a \emph{TLS client hello} 20.

Lastly, in the \emph{mitigation phase} (3), the SDN controller instructs the switches to block further connection attempts by the attackers and terminates the attacker's open connections to the target by injecting spoofed \texttt{RST} messages to the target system. This way, the target system recovers faster as it does not wait for the connections to actually time out. The process is repeated until the target system is reachable again and the observation phase cannot identify any additional attackers.

We have chosen an SDN-based architecture as it offers enough flexibility and extensibility to allow for an extensible system. Using SDN especially helps when setting up the identification phase and in the mitigation phase when blocking the attackers.

However, as yet, it is not capable of responding to potential slow attacks once detected at the targets. Therefore, in this work, we focus on identifying slow attackers and mitigating these attacks. Our framework does not require host-based protection methods and does not rely on support by the target host.

For this work, we only make minor, internal adjustments to the original framework. We focus primarily on the techniques used to classify clients as benign or malicious. One important change concerning the framework is the option to send \texttt{RST}-flagged TCP packets to the server under attack to deliberately terminate connections opened by attackers. Thus, when an attacker is identified, the resources occupied at the target can be freed almost instantly without waiting for the very long connection timeout of the server that is the very attack vector that gets exploited in the first place.


\subsection{Identification Schemes}
\label{schemes}

Identification of slow attackers is hindered by several factors. The attacking clients behave according to specification, the data rate of the attack is low, and\,---\,in case of a highly distributed attack\,---\,each client only opens a small amount of connections. This leads to intrusion detection systems not being able to successfully distinguish attacks from regular server traffic~\cite{apacheMit}. Currently, many servers such as Apache can be configured to mitigate the effect of slow HTTP attacks by reducing the maximum time a server waits to receive a full request. However, these changes also block legitimate requests from clients with slow Internet connections and an attack still has a noticeable impact on the server's performance~\cite{evalControls}. Moreover, this mitigation technique requires the administrator to become active and is therefore not a viable option for our use case.

Attacks conducted by the most common tools, however, also show common characteristics in terms of network traffic patterns. Based on this traffic (and extending on preliminary deliberations~\cite{tuebingen}), we identified six attacker identification schemes that we took into account for our evaluation:

\paragraph{Long Connections (LC)} A very basic method measures the duration $d$ of connections and deems very long connections suspicious. This method, however, needs to wait for the connection to last longer than a certain threshold in the area of minutes and, thus, the identification of attackers can take longer than with other methods. This could also lead to many false positives when the time out is set too short. This is the closest to the already established mitigation method of changing the aforementioned server settings.

\paragraph{Low packet rate (LPR)} One of the core characteristics of slow attackers is a low packet rate $p$ as the attackers try to send as little packets as possible while still keeping the connection open. This scheme alone might also lead to slow regular clients to be blocked. The packet rate is defined as the amount of packets divided by the elapsed time since the first packet of the connection.

\paragraph{Packet distance uniformity (PDU)} Even time intervals between packets are a feature that can be observed with scripted attacks. The assumption is that non-scripted real clients would send with varying packet rates due to the user behavior, network utilization, and available processing resources. Especially clients with bad connections that could be mistaken for attackers by the LPR metric experience differing packet distances. To reduce load, we only consider the packet distance for three consecutive packets in a row. The packet distance between two subsequent packets is defined as the difference between their receiving time stamps. The PDU is defined as the absolute value of the difference between the packet distances of three consecutive packets abbreviated as $\Delta$. 

\paragraph{Combination of LPR and PDU (LPR-PDU)} LPR can reliably detect slow clients while PDU can reliably detect constant clients while both mechanisms cannot assess the other trait. Therefore, a combination of both schemes could lead to better results. Both traits are present in attackers but should not be common in benign clients. The combination of the two metrics evaluates whether a client shows two characteristics typically attributed to an attacker. As a combination scheme, this scheme requires two thresholds $p$ and $\Delta$.

\paragraph{Low mean packet rate (MPR)} The mean packet rate $\bar p$ of an attack connection should be high compared to benign clients and could therefore also be used as an indicator for an attack.

\paragraph{Low packet rate variance (PRV)} The mean packet rate of an attack connection should be more consistent compared to benign clients because of the periodically generated traffic for keep alive. Therefore, we also analyze whether the packet rate variance $\sigma^2$ can be used as an indicator for attackers. \\

During our preliminary tests, we have noticed that these schemes (except for LC) behave differently whether the TCP handshake is taken into account. Therefore, we evaluate these with and without measuring the handshake packets.

The chosen schemes require very little calculation effort, work solely on the network layer with therefore comparatively minor privacy implications, and have very low storage requirements (at max, two values per client need to be stored). In addition to the aforementioned schemes, we also identify two schemes we choose not to evaluate further as they do not fulfill our requirements of a light-weight network-based scheme:

\paragraph{Incomplete application-layer messages} 
This scheme operates on the same conceptual level as the approach of Hong et al.~\cite{Hong2017} for detecting Slow HTTP attacks.
Benign clients typically do not send incomplete headers on application layer. However, incomplete headers are an inherent feature of slow attacks. Therefore, this method helps to identify attackers reliably. However, it relies heavily on the application layer protocol, a specific detector per protocol and attack type is necessary. For slow HTTP header attacks, for instance, the identification of incomplete packets needs to check \texttt{GET} requests for only one end-of-line character at the end or compare the \texttt{Content-length} definition with the actual body length of messages. This identification method is quite resource intensive as deep packet inspection is necessary and requires access to the application layer of the connections. In contrast to the other schemes, encrypted communication (e.g. TLS) cannot be analyzed.

\paragraph{Scoring-based Mechanism} A scoring-based system based on our prior work used to mitigate flooding attacks would rate every connection depending on the load caused at the target. For example, a scoring system would rate Slow POST attacks by giving a high score to packets belonging to a \texttt{POST} request. To prevent non-distributed DoS attacks, the number of connections per single clients can be considered. Thereby, any additional connection will increase the score which is assigned to a client. If other methods fail to identify attackers correctly, this mechanism can at least provide the support for a small subgroup of slow HTTP attacks (non-distributed attacks). Preliminary tests have shown that this scheme is very unreliable for slow attacks and is therefore excluded from the evaluation.


\subsection{Implementation}

The network monitor for the observation phase is based on Bro 2.5-372\footnote{\url{https://bro.org}}, which provides an easy to extend event-based system. Bro already offers some of the metrics necessary for the analysis such as the packet rate as built-in functions. Other metrics such as the packet rate without the TCP handshake have to be calculated without a built-in function of Bro. This can lead to more processing effort for these schemes as the built-in functions have been potentially optimized over the years while the newly added features might be less efficient.

Bro provides an API to the Ryu SDN controller\footnote{\url{https://osrg.github.io/ryu/}} called broccoli. As Ryu also satisfies our other requirements of extensibility, sufficient support of the OpenFlow protocol and our switches, we choose Ryu in version 4.6 for our prototype.

Attackers for the first attack scenario are simulated with the tools slowloris 0.1.4\footnote{\url{https://github.com/gkbrk/slowloris}} for slow header attacks 
and slowhttptest 1.6\footnote{\url{https://github.com/shekyan/slowhttptest}} for slow body attacks.

For the second attack scenario, we modify the original slowloris tool and create a variant with a less predictable behavior, called slowloris-ng\footnote{\url{https://github.com/vs-uulm/slowloris-ng}}.


\section{Evaluation}
\label{sec:evaluation}

The following chapter presents the evaluation results of the schemes introduced as part of the mitigation framework. We describe the setup and workloads based on real-life traffic scenarios.
We then present the results and discuss their implications.

\subsection{Setup}

Each scheme can be configured with specific parameters, for example, for the long connection scheme a threshold defines after how many seconds a connection is considered suspicious. For other schemes, such as low packet rate, the optimal rate needs to be determined. The optimal values are extracted in Section~\ref{results} by testing the behavior of the framework with a one-day traffic pcap, as described in Section~\ref{workloads}. The number of suspicious packets per client (number of strikes) necessary to deem a client an attacker is another parameter under investigation. Its purpose is to reduce false positives when a benign client sends only one packet or a very small amount of packets that incidentally fall below the threshold. However, this should have an impact on the detection time.


\subsection{Evaluation Workloads}
\label{workloads}

\begin{table}[]
\centering
\caption{Overview over the data sets used.}
\label{tab:sets}
\begin{tabular}{p{7em}p{7em}p{7em}p{7em}p{7em}}
\toprule
      & start date & duration & \#hosts & size (header~only) \\
\midrule
SUEE1 & 2017-11-02 & 24 h     & 1634      & 164 MB  \\
SUEE8 & 2017-11-05 & 8 d   & 8286      & 1871 MB \\
ICSI  & 2004-10-04 & 20 min   & 461       & 677 MB  \\
\bottomrule
\end{tabular}
\end{table}

We use three different evaluation workloads with benign traffic to test the mitigation system in terms of precision and accuracy (Table~\ref{tab:sets}). All test traces have been captured as actual live traffic from real deployments and merged with labeled attack traffic from the three attack tools under investigation.

For one, two recordings from the student union for electrical engineering at Ulm University\footnote{\url{https://fs-et.de}} have been used, one containing 24 hours (2nd to 3th November 2017 with 1,634 clients, SUEE1) and another containing eight days (5th to 13th November 2017 with 8,286 clients, SUEE8) of traffic data respectively. The web server of the student union offers information about the union on its main site, provides public real-time transport information for bus stops in the city which is used primarily on mobile devices via mobile networks, as well as several external and internal services. 

Additionally, we use a recording of an internal enterprise network of the Lawrence Berkeley National Laboratory / International Computer Science Institute (LBNL / ICSI) (from the 4th October 2004 between 10:03 pm and 10:23 pm with 461 clients, in the following called ICSI)\footnote{\url{https://www.icir.org/enterprise-tracing/download.html}}. 

All three pcap files contain only header data since the data sets were anonymized and do not contain application layer payload due to privacy concerns. There have been no attacks reported during the times of recording of the benign data sets. The data sets serve the following purpose in our analysis: SUEE1 is used as training data set to determine the best thresholds for each detection scheme. SUEE8 then is used to determine whether the trained mitigation system is capable to mitigate attacks adequately. The ICSI set is used as an indicator whether the parameters could hold universal value or if training for a specific network is necessary.

For the first attack scenario, the attacking tools are adapted to allow IP spoofing to simulate distributed attacks and are left in standard configuration apart from that. The parameters for slowhttptest are 30 seconds intervals, 8192 bytes for the Content-length header, 10 bytes POST-body length per packet and one socket per client. Slowloris is also configured to use only one socket per client. The default configuration is left in place in all other settings, resulting in a packet interval of 15 seconds.

For the second scenario, slowloris-ng includes several changes compared to the original slowloris. The additional features implement randomized behavior, which is configured to send in intervals of 15 seconds with a randomization interval of 5 seconds and sending the header lines as bursts of single messages per character. This tool shows how applicable the presented schemes are for highly improved attackers compared to easily accessible attacking scripts. For each of these tools, 49 to 50 clients are started simultaneously with different IP addresses to attack the web server. These attacks are run in parallel with the benign pcaps presented before. Due to the vast differences between the attacks, we choose to determine the best thresholds for each of the three attacks separately and evaluate these thresholds against all attacks.

We have combined the SUEE data sets with the attack recordings. We have published the data sets with a more detailed description on github\footnote{\url{https://github.com/vs-uulm/2017-SUEE-data-set}}. The MAC and IP addresses are anonymized, i.e. new addresses are set. Benign clients IP addresses in the anonymized data sets are moved to the 192.168.0.0/16 block, while attacking clients are in the 128.10.0.0/16 block.


\subsection{Results}
\label{results}

The evaluation thresholds are determined by testing the mitigation system with the SUEE1 data set with induced attacks of each type. The number of strikes for a detection is set to one. The thresholds are found using the bisection method starting with two extreme values that would result in detection of all clients\,---\,benign and attackers\,---\,and detection of no clients respectively. The balanced accuracy is used as quality metric. We have decided to use balanced accuracy over accuracy, as it takes the unbalance of the data sets into account (only 49 to 50 attackers versus up to 8,286 benign clients). Otherwise, a completely worthless classifier that classifies everything as benign would result in an accuracy of up to 0.994, while the balanced accuracy would be 0.5, similar to a coin toss, resulting in a much more accurate indicator. Compared to precision/recall and receiver operating characteristic (ROC), balanced accuracy has the advantage of resulting in one clear value, that can be taken as a good estimation of the quality of a scheme. Balanced accuracy (BACC) is defined as $BACC = (\frac{{TP}}{TP+FN} + \frac{TN}{TN+FP})\cdot~0.5$ with the true positive values TP, false positive FP, true negative TN, and false negative FN.

\begin{table}[]
\centering
\caption{Overview of the ideal thresholds for each scheme and attack for data set SUEE1.}
\label{tab:thresholds}
\begin{tabular}{lllll}
\toprule
Scheme  & \begin{tabular}[c]{@{}l@{}}TCP\\handshake\end{tabular} & slowloris       & slowhttptest   & slowloris-ng       \\
\midrule
  LC       & N/A     & $d=2.1\mathrm{e}{-5}$s      & $d=2.1\mathrm{e}{-5}$s     & $d=0.0999727$s     \\
\multirow{2}{*}{LPR}     & yes     & $p=0.091756$Hz       & $p=0.01739$Hz       & $p=0.783869$Hz        \\
        & no      & $p=0.079935$Hz        & $p=0.03806$Hz      & $p=0.77687$Hz  \\
\multirow{2}{*}{PDU}     & yes     & $\Delta=5.9\mathrm{e}{-5}$s       & $\Delta=2.5\mathrm{e}{-5}$s        & $\Delta=2.5\mathrm{e}{-5}$s  \\
        & no      & $\Delta=1.4\mathrm{e}{-5}$s       & $\Delta=0.000631$s     & $\Delta=1\mathrm{e}{-6}$s    \\
\multirow{2}{*}{LPR-PDU} & yes     & \begin{tabular}[c]{@{}l@{}}$p=0.091756$Hz\\ $\mathbf{\Delta=5.9\mathrm{e}{-5}}$\textbf{s}\end{tabular} & \begin{tabular}[c]{@{}l@{}}$p= 0.01739$Hz\\ $\Delta=2.5\mathrm{e}{-5}$s\end{tabular} & \begin{tabular}[c]{@{}l@{}}\textbf{$\mathbf{p=0.783869}$Hz}\\ $\Delta=4.1\mathrm{e}{-5}$s\end{tabular} \\
        & no      & \begin{tabular}[c]{@{}l@{}}$p=0.079935$Hz\\ $\Delta=1.4\mathrm{e}{-5}$s\end{tabular} & \begin{tabular}[c]{@{}l@{}}$p=0.03806$Hz\\ $\mathbf{\Delta=0.000631}$\textbf{s}\end{tabular}  & \begin{tabular}[c]{@{}l@{}}\textbf{$\mathbf{p=0.77687}$Hz}\\ $\Delta=1\mathrm{e}{-6}$s\end{tabular}      \\
\multirow{2}{*}{MPR}     & yes     & $\bar p=0.83315$Hz        & $\bar p=0.83315$Hz        & $\bar p=0.83315$Hz  \\
        & no      & $\bar p=4049$Hz     & $\bar p=21845$Hz   & $\bar p=995$Hz      \\
\multirow{2}{*}{PRV}     & yes     & $\sigma^2 = 0.028007\mathrm{Hz^2}$        & $\sigma^2 = 0.028007\mathrm{Hz^2}$       & $\sigma^2 = 0.028007\mathrm{Hz^2}$        \\
        & no      & $\sigma^2 = 1332497506\mathrm{Hz^2}$        & $\sigma^2 = 1332497506\mathrm{Hz^2}$     & $\sigma^2 = 1332497506\mathrm{Hz^2}$ \\      
\bottomrule
\end{tabular}
\end{table}

\begin{table}[]
\centering
\caption{Evaluation results (first part) dependent on scheme, whether or not TCP handshake is evaluated, data set, attack (SL: slowloris, SH: slowhttptest, NG: slowloris-ng) and thresholds; detection time is the mean and standard deviation of the detection times of the true positives in seconds.}
\label{tab:eval1}
\begin{tabular}{|l|l|l|l|p{3.5em}|p{3.5em}|p{3.5em}|p{3.5em}|p{3.5em}|l|}
\hline
\rotatebox[origin=l]{90}{\textbf{schemes}} & \rotatebox[origin=l]{90}{\textbf{TCP handshake}} & \rotatebox[origin=l]{90}{\textbf{data set}} & \rotatebox[origin=l]{90}{\textbf{attack}} & \rotatebox[origin=l]{90}{\textbf{true positive}} & \rotatebox[origin=l]{90}{\textbf{false positive}} & \rotatebox[origin=l]{90}{\textbf{false negative}} & \rotatebox[origin=l]{90}{\textbf{true negative}} & \rotatebox[origin=l]{90}{\textbf{balanced accuracy}} & \rotatebox[origin=l]{90}{\textbf{detection time t in s}}\\
\hline
\multirow{6}{*}{LC}       &   & \multirow{3}{*}{SUEE8} & SL & 32 & 4959 & 17 & 3544 & 0.535 & $\bar t=0.84~\sigma=1.35$  \\   
  &   &       & SH & 49 & 4959 & 0  & 3544 & 0.708 & $\bar t=12.86~\sigma=8.81$  \\   
  &   &       & NG & 50 & 8502 & 0  & 1    & 0.5   & $\bar t=0.12~\sigma=0.59$  \\ \cline{3-10} 
  &   & \multirow{3}{*}{ICSI}  & SL & 32 & 19   & 17 & 544  & 0.81  & $\bar t=0.85~\sigma=1.35$ \\   
  &   &       & SH & 49 & 19   & 0  & 544  & 0.983 & $\bar t=12.86~\sigma=8.81$  \\  
  &   &       & NG & 50 & 562  & 0  & 1    & 0.501 & $\bar t=0.12~\sigma=0.59$  \\ \hline
\multirow{12}{*}{MPR}     & \multirow{6}{*}{N} & \multirow{3}{*}{SUEE8} & SL & 49 & 7690 & 0  & 813  & 0.548 & $\bar t=2.61~\sigma=5.10$\\  
  &   &       & SH & 49 & 7690 & 0  & 813  & 0.548 & $\bar t=24.34~\sigma=37.56$ \\  
  &   &       & NG & 50 & 7690 & 0  & 813  & 0.548 & $\bar t=4.39~\sigma=6.68$  \\ \cline{3-10} 
  &   & \multirow{3}{*}{ICSI}  & SL & 49 & 532  & 0  & 31   & 0.528 & $\bar t=2.61~\sigma=5.10$  \\  
  &   &       & SH & 49 & 532  & 0  & 31   & 0.528 & $\bar t=24.33~\sigma=37.56$ \\  
  &   &       & NG & 50 & 532  & 0  & 31   & 0.528 & $\bar t=4.39~\sigma=6.68$  \\ \cline{2-10} 
  & \multirow{6}{*}{Y} & \multirow{3}{*}{SUEE8} & SL & 19 & 611  & 30 & 7892 & 0.658 & $\bar t=71.81~\sigma=42.96$ \\  
  &   &       & SH & 49 & 611  & 0  & 7892 & 0.964 & $\bar t=108.73~\sigma=52.65$ \\  
  &   &       & NG & 14 & 611  & 36 & 7892 & 0.604 & $\bar t=106.34~\sigma=85.91$ \\ \cline{3-10} 
  &   & \multirow{3}{*}{ICSI}  & SL & 19 & 153  & 30 & 410  & 0.558 & $\bar t=71.81~\sigma=42.96$ \\  
  &   &       & SH & 49 & 153  & 0  & 410  & 0.864 & $\bar t=108.73~\sigma=52.65$ \\  
  &   &       & NG & 14 & 153  & 36 & 410  & 0.504 & $\bar t=106.34~\sigma=85.91$ \\ \hline
\multirow{12}{*}{PRV}     & \multirow{6}{*}{N} & \multirow{3}{*}{SUEE8} & SL & 49 & 7691 & 0  & 812  & 0.548 & $\bar t=1.07~\sigma=1.36$  \\  
  &   &       & SH & 49 & 7691 & 0  & 812  & 0.548 & $\bar t=20.58~\sigma=38.77$ \\  
  &   &       & NG & 50 & 7691 & 0  & 812  & 0.548 & $\bar t=1.60~\sigma=1.50$  \\ \cline{3-10} 
  &   & \multirow{3}{*}{ICSI}  & SL & 49 & 520  & 0  & 43   & 0.538 & $\bar t=1.07~\sigma=1.36$  \\  
  &   &       & SH & 49 & 520  & 0  & 43   & 0.538 & $\bar t=20.58~\sigma=38.77$ \\  
  &   &       & NG & 50 & 520  & 0  & 43   & 0.538 & $\bar t=1.60~\sigma=1.50$  \\ \cline{2-10} 
  & \multirow{6}{*}{Y} & \multirow{3}{*}{SUEE8} & SL & 24 & 1431 & 25 & 7072 & 0.661 & $\bar t=1.46~\sigma=1.19$  \\  
  &   &       & SH & 49 & 1431 & 0  & 7072 & 0.916 & $\bar t=82.18~\sigma=44.05$ \\  
  &   &       & NG & 13 & 1431 & 37 & 7072 & 0.546 & $\bar t=2.00~\sigma=0.00$  \\ \cline{3-10} 
  &   & \multirow{3}{*}{ICSI}  & SL & 24 & 70   & 25 & 493  & 0.683 & $\bar t=1.46~\sigma=1.19$  \\  
  &   &       & SH & 49 & 70   & 0  & 493  & 0.938 & $\bar t=82.18~\sigma=44.05$ \\  
  &   &       & NG & 13 & 70   & 37 & 493  & 0.568 & $\bar t=2.00~\sigma=0.00$  \\ \hline
\end{tabular}
\end{table}

\begin{table}[]
\centering
\caption{Evaluation results (second part) dependent on scheme, whether or not TCP handshake is evaluated, data set, attack (SL: slowloris, SH: slowhttptest, NG: slowloris-ng) and thresholds; detection time is the mean and standard deviation of the detection times of the true positives in seconds.}
\label{tab:eval2}
\begin{tabular}{|l|l|l|l|p{3.5em}|p{3.5em}|p{3.5em}|p{3.5em}|p{3.5em}|l|}
\hline
\rotatebox[origin=l]{90}{\textbf{schemes}} & \rotatebox[origin=l]{90}{\textbf{TCP handshake}} & \rotatebox[origin=l]{90}{\textbf{data set}} & \rotatebox[origin=l]{90}{\textbf{attack}} & \rotatebox[origin=l]{90}{\textbf{true positive}} & \rotatebox[origin=l]{90}{\textbf{false positive}} & \rotatebox[origin=l]{90}{\textbf{false negative}} & \rotatebox[origin=l]{90}{\textbf{true negative}} & \rotatebox[origin=l]{90}{\textbf{balanced accuracy}} & \rotatebox[origin=l]{90}{\textbf{detection time t in s}}\\
\hline
\multirow{12}{*}{LPR}     & \multirow{6}{*}{N} & \multirow{3}{*}{SUEE8} & SL   & 49 & 641  & 0  & 7862 & 0.962 & $\bar t=211.26~\sigma=28.65$ \\  
  &   &       & SH       & 49 & 403  & 0  & 8100 & 0.976 & $\bar t=210.21~\sigma=28.65$  \\  
  &   &       & NG        & 50 & 3853 & 0  & 4650 & 0.773 & $\bar t=52.87~\sigma=51.39$  \\ \cline{3-10} 
  &   & \multirow{3}{*}{ICSI}  & SL       & 49 & 165  & 0  & 398  & 0.853 & $\bar t=211.25~\sigma=28.65$\\  
  &   &       & SH        & 49 & 107  & 0  & 456  & 0.905 & $\bar t=210.21~\sigma=0.01$ \\  
  &   &       & NG        & 50 & 336  & 0  & 227  & 0.702 & $\bar t=52.87~\sigma=51.39$   \\ \cline{2-10} 
  & \multirow{6}{*}{Y} & \multirow{3}{*}{SUEE8} & SL      & 49 & 1019 & 0  & 7484 & 0.94  & $\bar t=174.83~\sigma=34.78$ \\  
  &   &       & SH        & 49 & 139  & 0  & 8364 & 0.992 & $\bar t=240.06~\sigma=0.07$  \\  
  &   &       & NG       & 50 & 4242 & 0  & 4261 & 0.751 & $\bar t=38.77~\sigma=55.31$  \\ \cline{3-10} 
  &   & \multirow{3}{*}{ICSI}  & SL       & 49 & 185  & 0  & 378  & 0.836 & $\bar t=174.84~\sigma=34.78$  \\  
  &   &       & SH        & 49 & 64   & 0  & 499  & 0.943 & $\bar t=240.06~\sigma=0.07$  \\  
  &   &       & NG       & 50 & 349  & 0  & 214  & 0.69  & $\bar t=38.77~\sigma=55.31$ \\ \hline
\multirow{12}{*}{PDU}     & \multirow{6}{*}{N} & \multirow{3}{*}{SUEE8} & SL       & 49 & 1884 & 0  & 6619 & 0.889 & $\bar t=46.09~\sigma=37.79$ \\  
  &   &       & SH       & 49 & 3502 & 0  & 5001 & 0.794 & $\bar t=105.63~\sigma=42.73$ \\  
  &   &       & NG       & 50 & 538  & 0  & 7965 & 0.968 & $\bar t=12.22~\sigma=14.01$  \\ \cline{3-10} 
  &   & \multirow{3}{*}{ICSI}  & SL       & 49 & 278  & 0  & 285  & 0.753 & $\bar t=46.09~\sigma=37.79$ \\  
  &   &       & SH       & 49 & 396  & 0  & 167  & 0.648 & $\bar t=105.63~\sigma=42.73$ \\  
  &   &       & NG       & 50 & 212  & 0  & 351  & 0.812 & $\bar t=12.21~\sigma=14.01$  \\ \cline{2-10} 
  & \multirow{6}{*}{Y} & \multirow{3}{*}{SUEE8} & SL       & 49 & 4021 & 0  & 4482 & 0.764 & $\bar t=5.94~\sigma=33.78$  \\  
  &   &       & SH       & 49 & 3407 & 0  & 5096 & 0.8   & $\bar t=5.88~\sigma=6.69$    \\  
  &   &       & NG       & 49 & 3407 & 1  & 5096 & 0.79  & $\bar t=1.56~\sigma=1.49$  \\ \cline{3-10} 
  &   & \multirow{3}{*}{ICSI}  & SL       & 49 & 380  & 0  & 183  & 0.663 & $\bar t=5.94~\sigma=33.78$ \\  
  &   &       & SH       & 49 & 352  & 0  & 211  & 0.687 & $\bar t=5.88~\sigma=6.69$\\  
  &   &       & NG        & 49 & 352  & 1  & 211  & 0.677 & $\bar t=1.56~\sigma=1.49$   \\ \hline
\multirow{12}{*}{\begin{tabular}[c]{@{}l@{}}LPR-\\PDU\end{tabular}} & \multirow{6}{*}{N} & \multirow{3}{*}{SUEE8} & SL & 49 & 217  & 0  & 8286 & 0.987 & $\bar t=211.26~\sigma=28.65$ \\
  &   &       & SH  & 49 & 197  & 0  & 8306 & 0.988 & $\bar t=210.21~\sigma=0.01$  \\  
  &   &       & NG     & 50 & 315  & 0  & 8188 & 0.981 & $\bar t=55.85~\sigma=50.06$ \\ \cline{3-10} 
  &   & \multirow{3}{*}{ICSI}  & SL  & 49 & 102  & 0  & 461  & 0.909 & $\bar t=211.26~\sigma=28.65$ \\  
  &   &       & SH  & 49 & 86   & 0  & 477  & 0.924 & $\bar t=210.21~\sigma=0.01$  \\  
  &   &       & NG     & 50 & 164  & 0  & 399  & 0.854 & $\bar t=55.85~\sigma=50.06$ \\ \cline{2-10} 
  & \multirow{6}{*}{Y} & \multirow{3}{*}{SUEE8} & SL  & 49 & 471  & 0  & 8032 & 0.972 & $\bar t=176.36~\sigma=35.93$ \\  
  &   &       & SH   & 49 & 88   & 0  & 8415 & 0.995 & $\bar t=240.06~\sigma=0.07$  \\  
  &   &       & NG & 50 & 1509 & 0  & 6994 & 0.911 & $\bar t=39.21~\sigma=55.03$ \\ \cline{3-10} 
  &   & \multirow{3}{*}{ICSI}  & SL & 49 & 134  & 0  & 429  & 0.881 & $\bar t=176.36~\sigma=35.93$ \\  
  &   &       & SH & 49 & 33   & 0  & 530  & 0.971 & $\bar t=240.01~\sigma=0.07$  \\  
  &   &       & NG  & 50 & 270  & 0  & 293  & 0.76  & $\bar t=39.21~\sigma=55.03$  \\ 
\hline
\end{tabular}
\end{table}

\begin{table}[]
\centering
\caption{Evaluation results for LPR-PDU when for each partial scheme the maximum threshold is chosen (bold values in Table~\ref{tab:thresholds}).}
\label{tab:eval3}
\begin{tabular}{|l|l|l|p{3.5em}|p{3.5em}|p{3.5em}|p{3.5em}|p{3.5em}|l|}
\hline
 \rotatebox[origin=l]{90}{\textbf{TCP handshake}} & \rotatebox[origin=l]{90}{\textbf{data set}} & \rotatebox[origin=l]{90}{\textbf{attack}} & \rotatebox[origin=l]{90}{\textbf{true positive}} & \rotatebox[origin=l]{90}{\textbf{false positive}} & \rotatebox[origin=l]{90}{\textbf{false negative}} & \rotatebox[origin=l]{90}{\textbf{true negative}} & \rotatebox[origin=l]{90}{\textbf{balanced accuracy}} & \rotatebox[origin=l]{90}{\textbf{detection time t in s}}\\
\hline
\multirow{6}{*}{N} & \multirow{3}{*}{SUEE8} & SL & 49 & 1261 & 0 & 7242 & 0.926 & $\bar t=20.35~\sigma=11.28$ \\ 
                   &                        & SH  & 49 & 1261 & 0 & 7242 & 0.926 & $\bar t=107.47~\sigma=38.59$ \\ 
                   &                        & NG        & 50 & 1261 & 0 & 7242 & 0.926 & $\bar t=52.87~\sigma=51.39$ \\ \cline{2-9} 
                   & \multirow{3}{*}{ICSI}  & SL & 49 & 277  & 0 & 286  & 0.754 & $\bar t=20.35~\sigma=11.28$ \\ 
                   &                        & SH  & 49 & 277  & 0 & 286  & 0.754 & $\bar t=107.47~\sigma=38.59$ \\ 
                   &                        & NG        & 50 & 277  & 0 & 286  & 0.754 & $\bar t=52.87~\sigma=51.39$ \\ \hline
\multirow{6}{*}{Y} & \multirow{3}{*}{SUEE8} & SL & 49 & 1603 & 0 & 6900 & 0.906 & $\bar t=14.83~\sigma=33.02$ \\ 
                   &                        & SH  & 49 & 1603 & 0 & 6900 & 0.906 & $\bar t=13.05~\sigma=7.75$ \\ 
                   &                        & NG        & 50 & 1603 & 0 & 6900 & 0.906 & $\bar t=39.21~\sigma=55.03$ \\ \cline{2-9} 
                   & \multirow{3}{*}{ICSI}  & SL & 49 & 272  & 0 & 291  & 0.758 & $\bar t=14.83~\sigma=33.02$ \\ 
                   &                        & SH  & 49 & 272  & 0 & 291  & 0.758 & $\bar t=13.05~\sigma=7.75$ \\ 
                   &                        & NG        & 50 & 272  & 0 & 291  & 0.758 & $\bar t=39.21~\sigma=55.03$ \\ \hline
\end{tabular}
\end{table}




The thresholds determined by this test can be seen in Table~\ref{tab:thresholds}. The table shows, that some schemes are very similar for all attacks (e.g. MPR, PRV; LC for slowloris and slowhttptest) while other schemes show big differences for different attacks (e.g. LPR-PDU). MPR and PRV show extreme differences depending on if the TCP handshake is part of the evaluation or not. The high thresholds when ignoring the handshake might imply, that these schemes might not be applicable without the TCP handshake. For LPR-PDU, in addition to the best thresholds for each attack, we also evaluate the maximum values for each partial scheme (highlighted in the table in bold).

For every classification scheme, there are two things to consider. On the one hand, how precise the identification is of each scheme for each attack, measuring if the attack can be mitigated successfully without too many blocked benign clients. We again use the balanced accuracy to assess the quality of the scheme but report all false/true positive and false/true negative values as well. Furthermore, another very important aspect is the detection time, i.e. the mean time each attacker remains undetected. Slow attacks work by opening as many connections as possible and keeping them open as long as possible to ensure maximum impact. If all attackers can be identified correctly but the detection time is too high, the attack might still be successful. It might even be worth trading accuracy for lower detection times if necessary.

Table~\ref{tab:eval1} shows the results for the long connection scheme (LC), the low mean packet rate scheme (MPR) and the low packet rate variance scheme (PRV). LC can be used to detect slowhttptest fast with a highly varying false positive rate between 3.4\% and 58\%. It cannot be used to detect the other attacks. For slowloris-ng, it behaves on the same level as a coin toss.
MPR and PRV show similar results. They can detect slowhttptest but are close to useless for the other attacks. For these three schemes, the TCP handshake has to be taken into account.

Table~\ref{tab:eval2} contains our results for the schemes low packet rate (LPR), packet distance uniformity (PDU), and the combination of both schemes. The results show, that all metrics can be used to detect attacks (except for rare cases, all attackers were found), however, the false positive rate varies significantly.

Low packet rate is a good classifier for the basic attacks slowloris and slowhttptest with a balanced accuracy of 0.96 to 0.98 for the SUEE8 data set. However, detection times of up to 210 seconds per client have to be considered.
Packet distance uniformity is much faster but also less reliable than LPR for the basic attacks, it performs better than PTR when faced with the improved slowloris-ng attack.
The combination of the aforementioned schemes shows much better results than the two schemes each alone. With a balanced accuracy between 0.854 (untrained data set) and 0.987 (trained data set) without TCP handshake and 0.76 (untrained data set) to 0.995 (trained data set) with TCP handshake, this method proves to be the most reliable scheme. However, as a combined method, it also inherits the high detection time of LPR with the best threshold pairs for each attack. 

Up to here, we evaluated whether the schemes can work with the right thresholds for each attack. However, when defending a real network we do not know which attack the attacker will choose. For the most promising schemes (LPR, PDU, and LPR-PDU) we therefore also evaluated how these schemes hold up when the threshold is not the ideal one for these attacks.

\begin{figure}
  \centering
  \parbox{\textwidth}{

  \centering
    \hspace{1.3em}
    (TCP handshakes excluded)
    \hspace{6em}(TCP handshakes included)
  }

  \centering
  Low packet rate (LPR).\\
  \parbox{0.495\textwidth}{
    \includegraphics[width=0.5\textwidth]{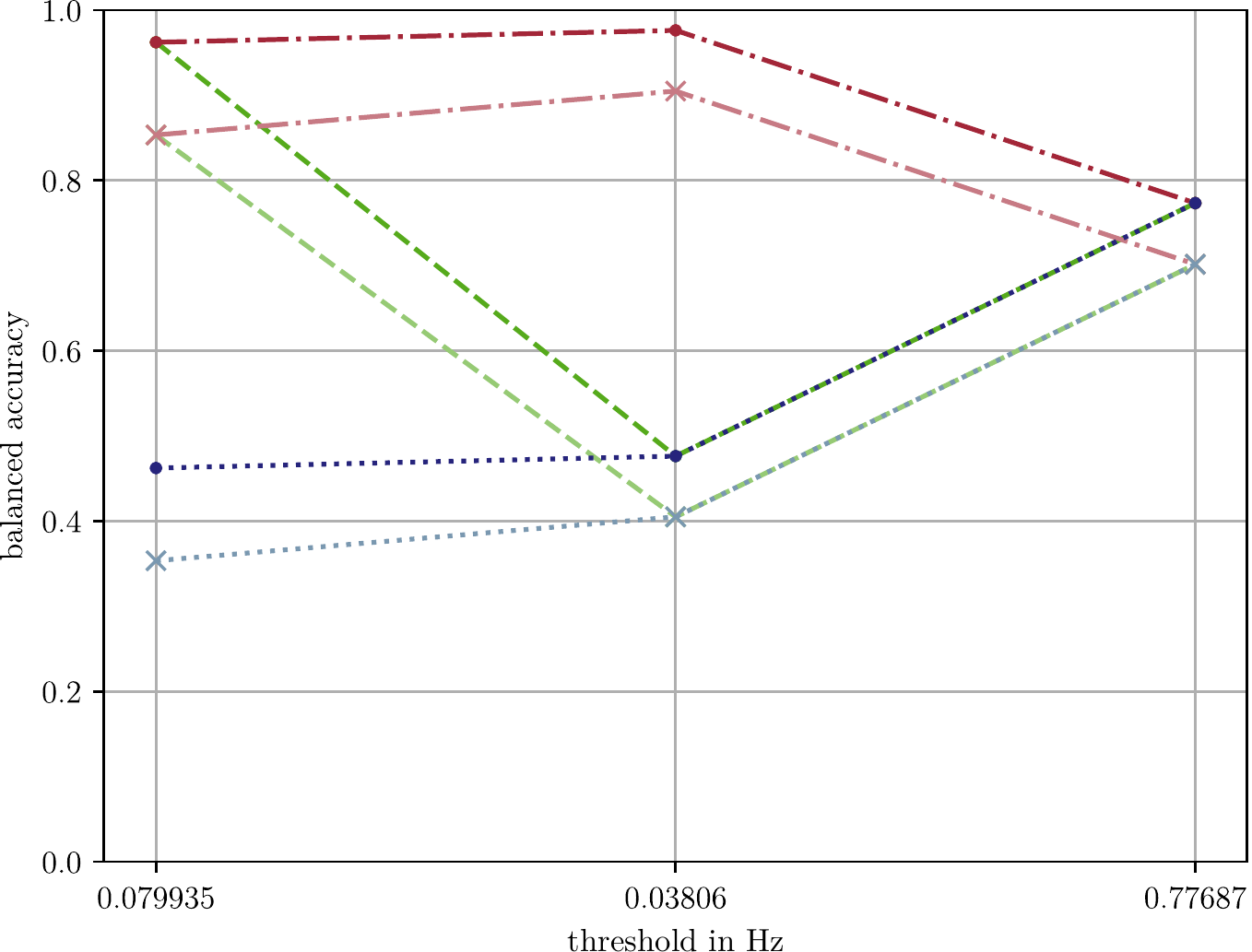}
  }
  \hfill
  \parbox{0.495\textwidth}{
    \includegraphics[width=0.5\textwidth]{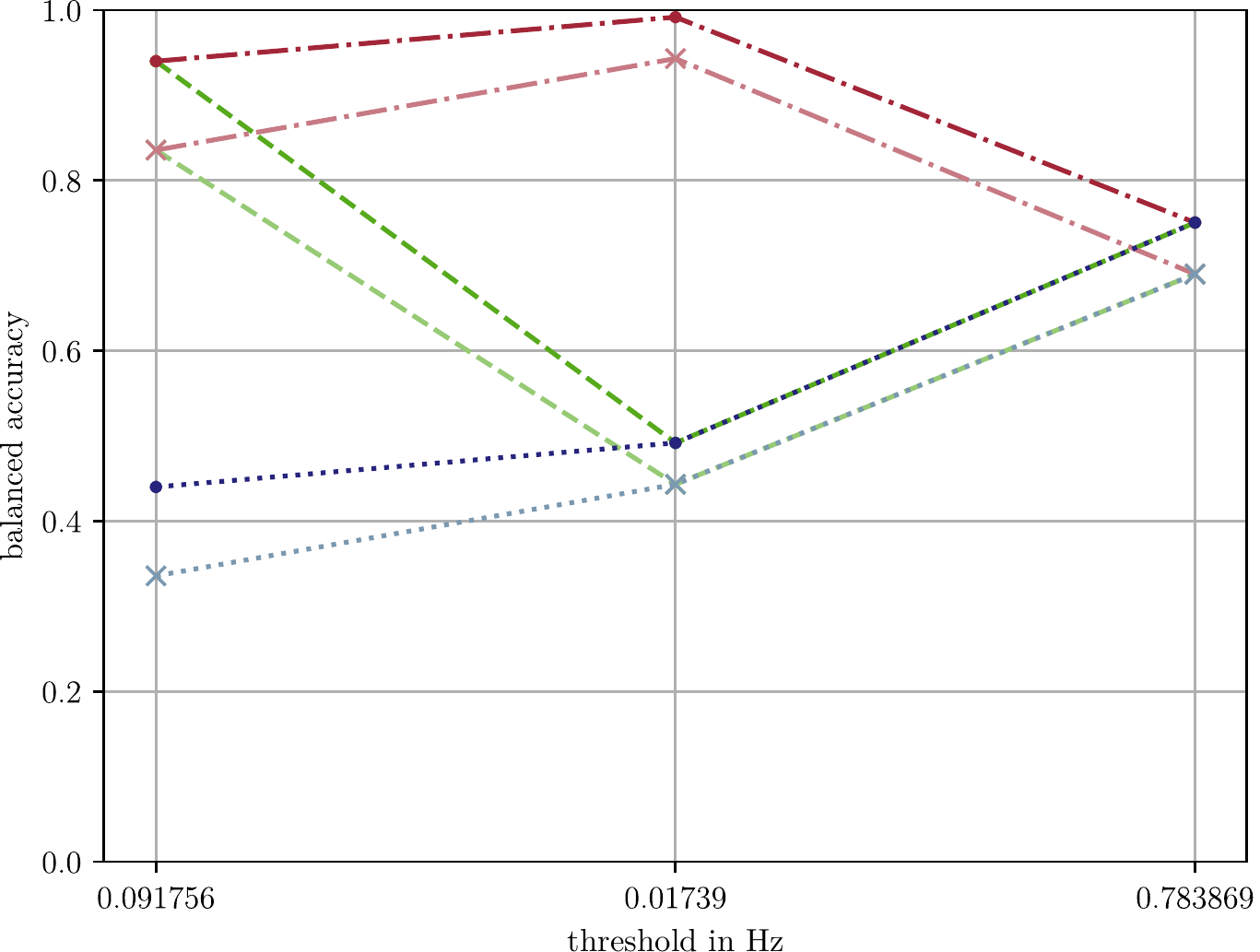}
  }

  \vspace{1em}
\centering
  Packet distance uniformity (PDU).\\
  \parbox{0.495\textwidth}{
    \includegraphics[width=0.5\textwidth]{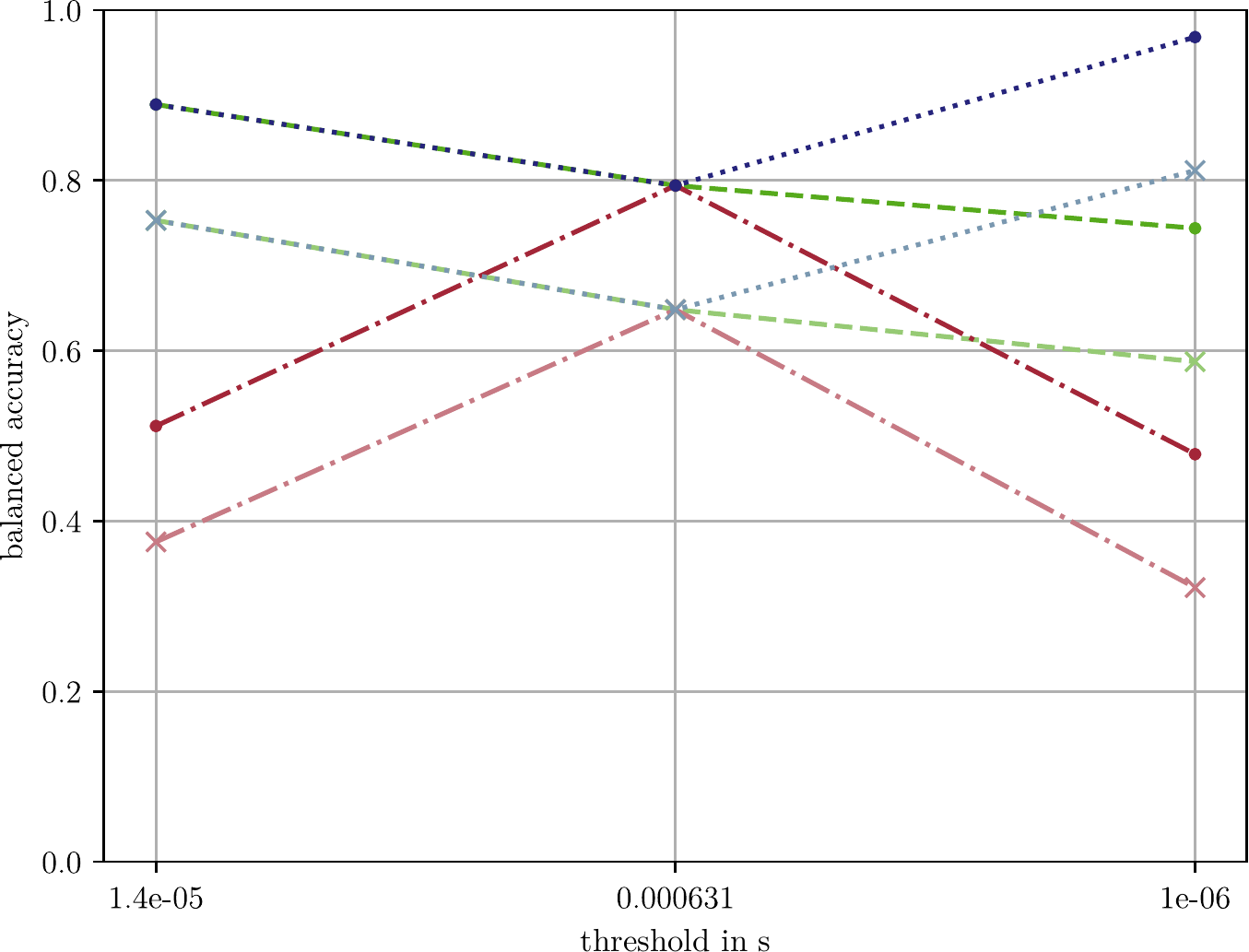}
  }
  \hfill
  \parbox{0.495\textwidth}{
    \includegraphics[width=0.5\textwidth]{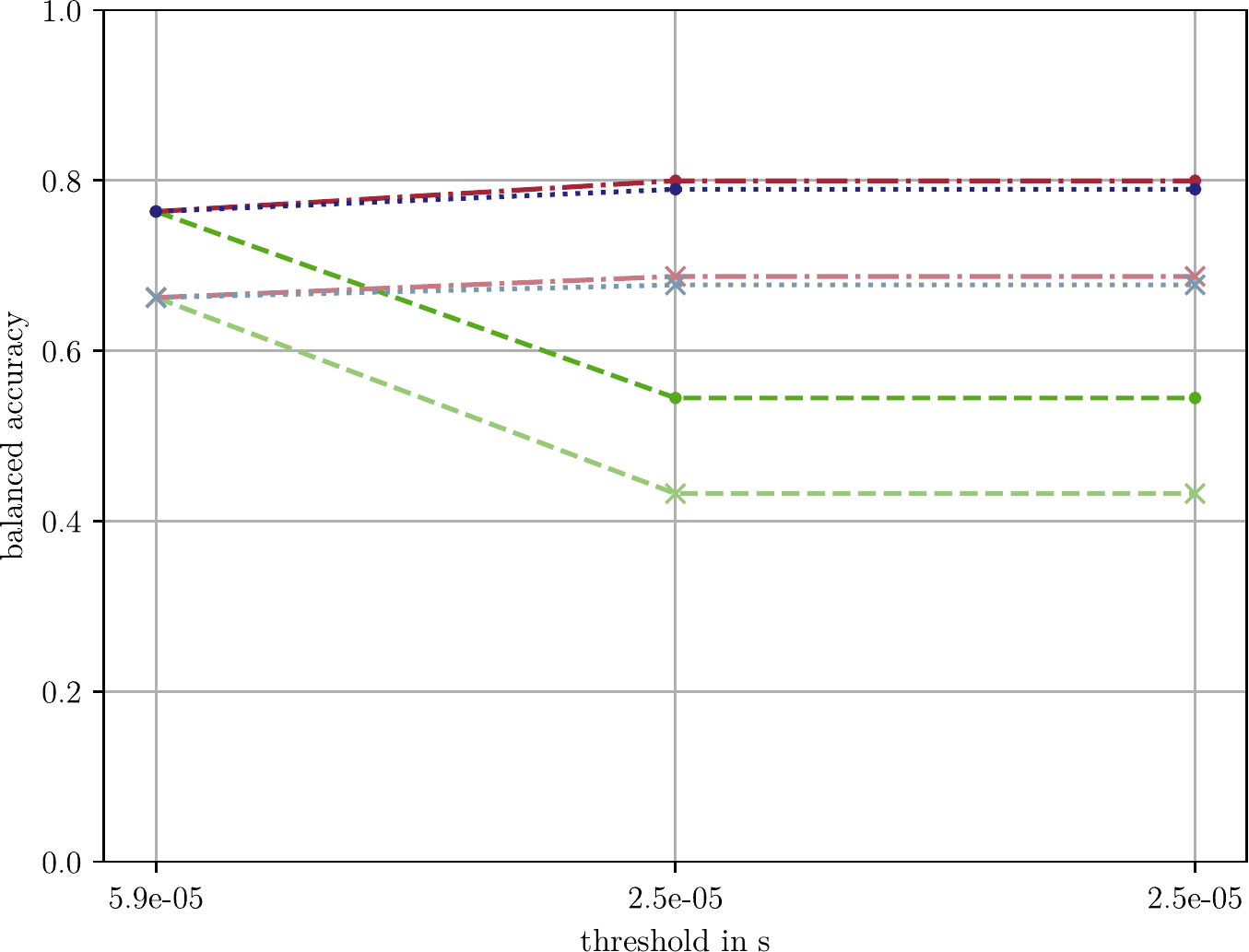}
  }

  \vspace{1em}
\centering
  Combination of low packet rate and packet distance uniformity (LPR-PDU).
  \parbox{0.495\textwidth}{
    \includegraphics[width=0.5\textwidth]{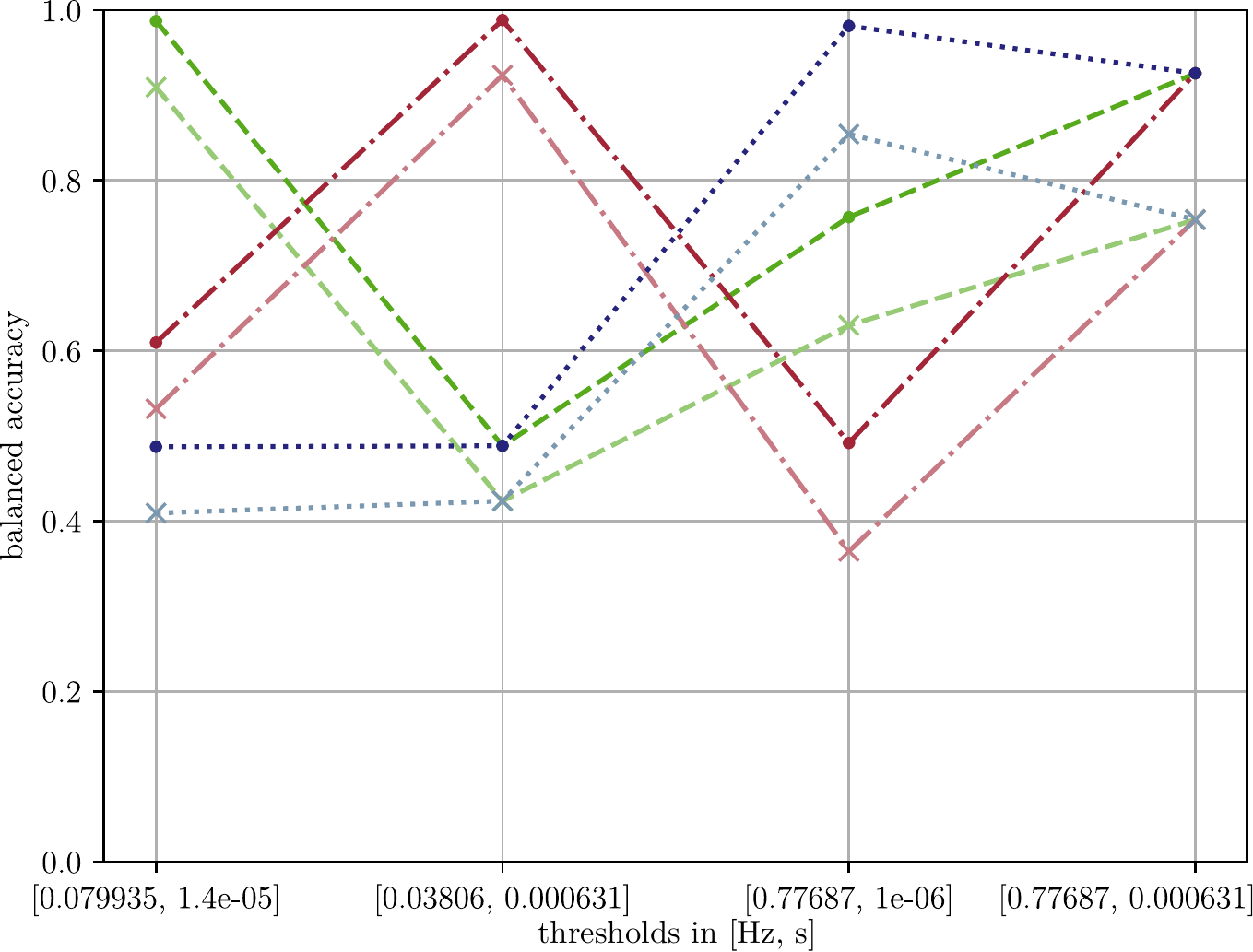}
  }
  \hfill
  \parbox{0.495\textwidth}{
    \includegraphics[width=0.5\textwidth]{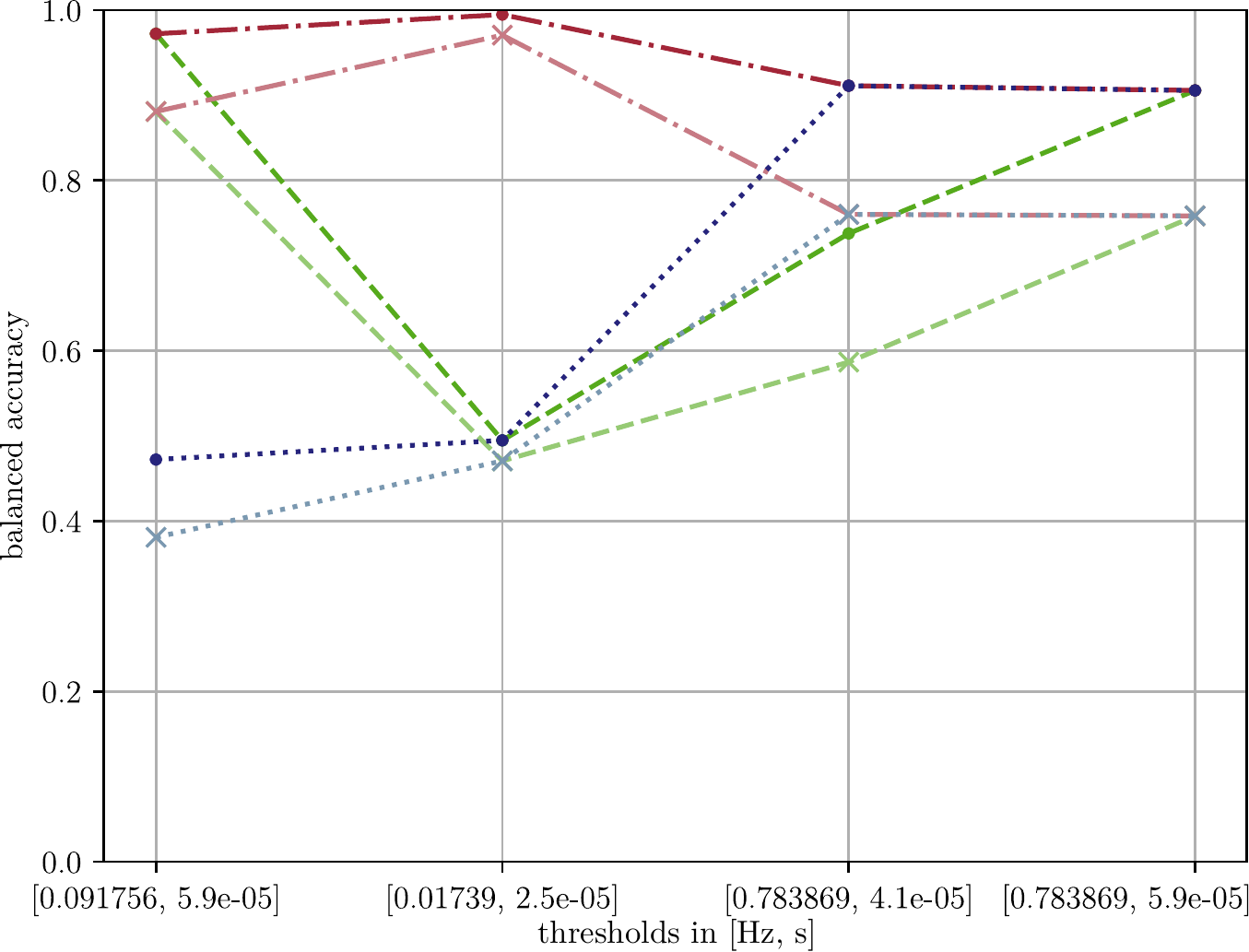}
  }



  \parbox{0.3\textwidth}{
  }
  \hfill
  \parbox{0.4\textwidth}{
    \includegraphics[width=0.4\textwidth]{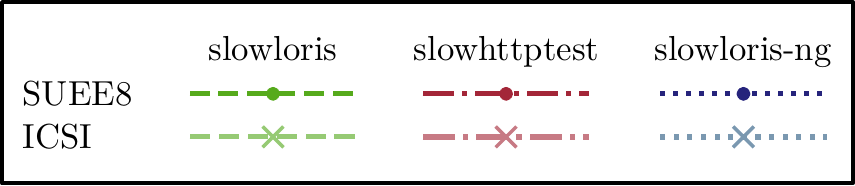}
  }
  \hfill  
  \parbox{0.3\textwidth}{
  }

  \caption{Balanced accuracy for LPR, PDU, and LPR-PDU (top to bottom) without TCP handshake and with TCP handshake (left to right). Each with the best thresholds for slowloris, slowhttptest, and slowloris-ng (left to right) and for LPR-PDU additionally the results when for each partial scheme threshold the maximum value of the three is used.}\label{fig:bacc-graphs}
   \label{fig:bacc}
\end{figure}

Figure~\ref{fig:bacc} shows extended balanced accuracy results for LPR, PDU, and LPR-PDU. For each threshold (or threshold pair) the diagrams show how good the classifiers are identifying the attackers in all three attacks. 
For LPR in the top two graphs, it can be seen that the best threshold for slowloris also works well against slowhttptest (with the same detection rate). The best threshold for slowhttptest, however, is unusable both for slowloris and slowloris-ng. As slowloris-ng has a higher packet rate than the other two attacks, their ideal threshold is much higher. This means more false positives and therefore a lower balanced accuracy for all schemes. However, all attackers were detected reliably with this threshold.

For the packet distance uniformity scheme, when not taking the TCP handshake into account, the best threshold for slowloris also shows the same results for slowloris-ng. However, for slowhttptest, this value is unusable. In turn, the best threshold for slowloris-ng is unusable for the other two schemes. A packet distance of $0.000631s$ as the highest value results in more false positives than the other values but results in a true positive rate of 100\% for all attacks. When we include the TCP handshake, slowhttptest and slowloris-ng work best with the same threshold which is not usable for slowloris while the slowloris threshold results in a high false positive rate for all attacks.

The combined scheme is evaluated with four different threshold pairs: The best pair for each attack and the maximum threshold values for each of these pairs. The best thresholds for the three attacks show, that this scheme works very well in detecting each attack (with a balanced accuracy of up to 0.995). However, each of these threshold pairs results in very bad detection rates for the other attacks. Combining the maximum thresholds of each partial scheme, however, results in a good balanced accuracy of up to 0.926 without TCP handshake or 0.906 with TCP handshake for all attacks equally. The detection times of 13 to 39 seconds also are much better than the ideal thresholds for each attack.

\begin{figure}
\centering
  \parbox{0.495\textwidth}{
    
    \includegraphics[width=0.5\textwidth]{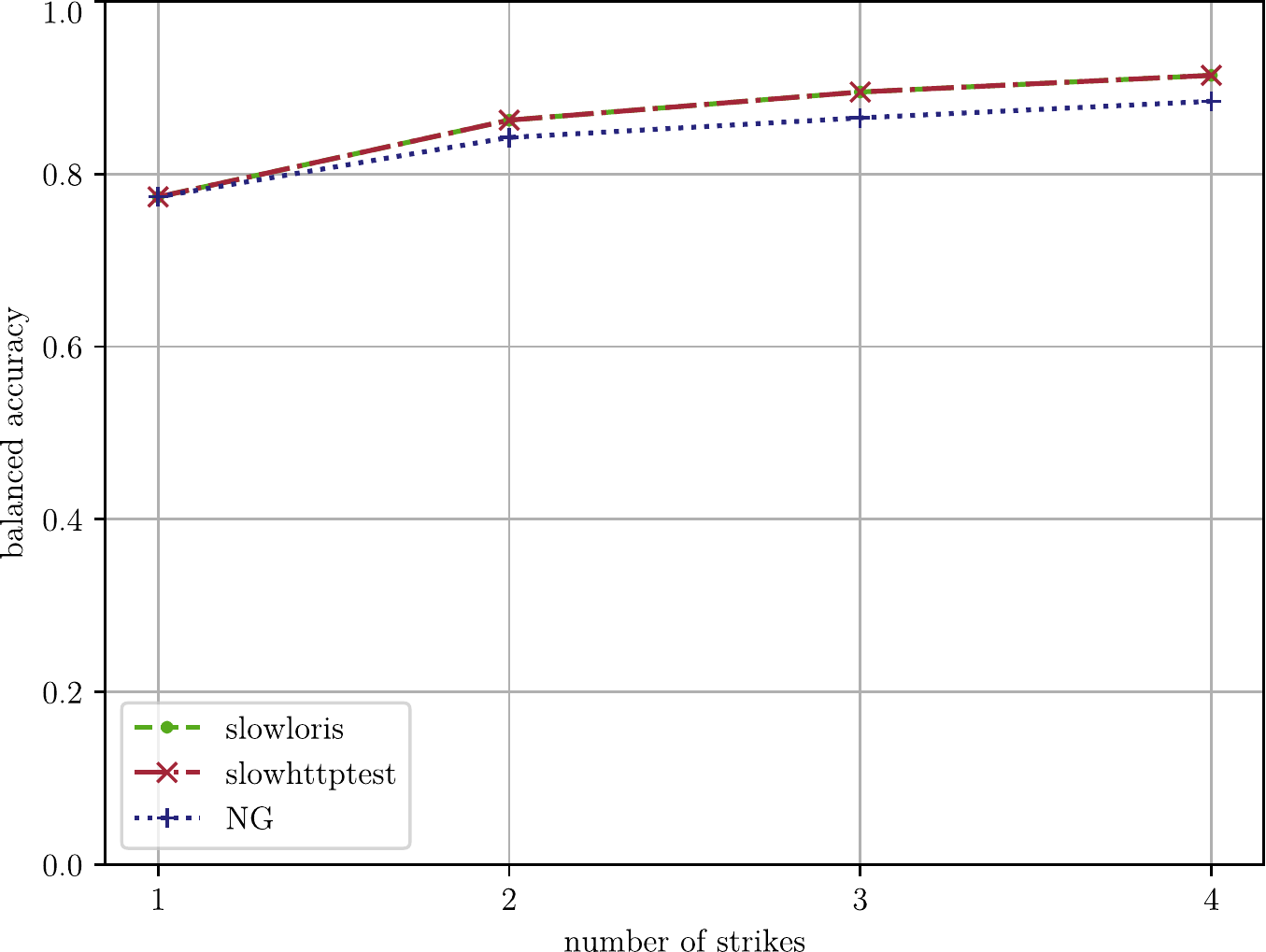}
  }
  \hfill
  \parbox{0.495\textwidth}{
    \includegraphics[width=0.5\textwidth]{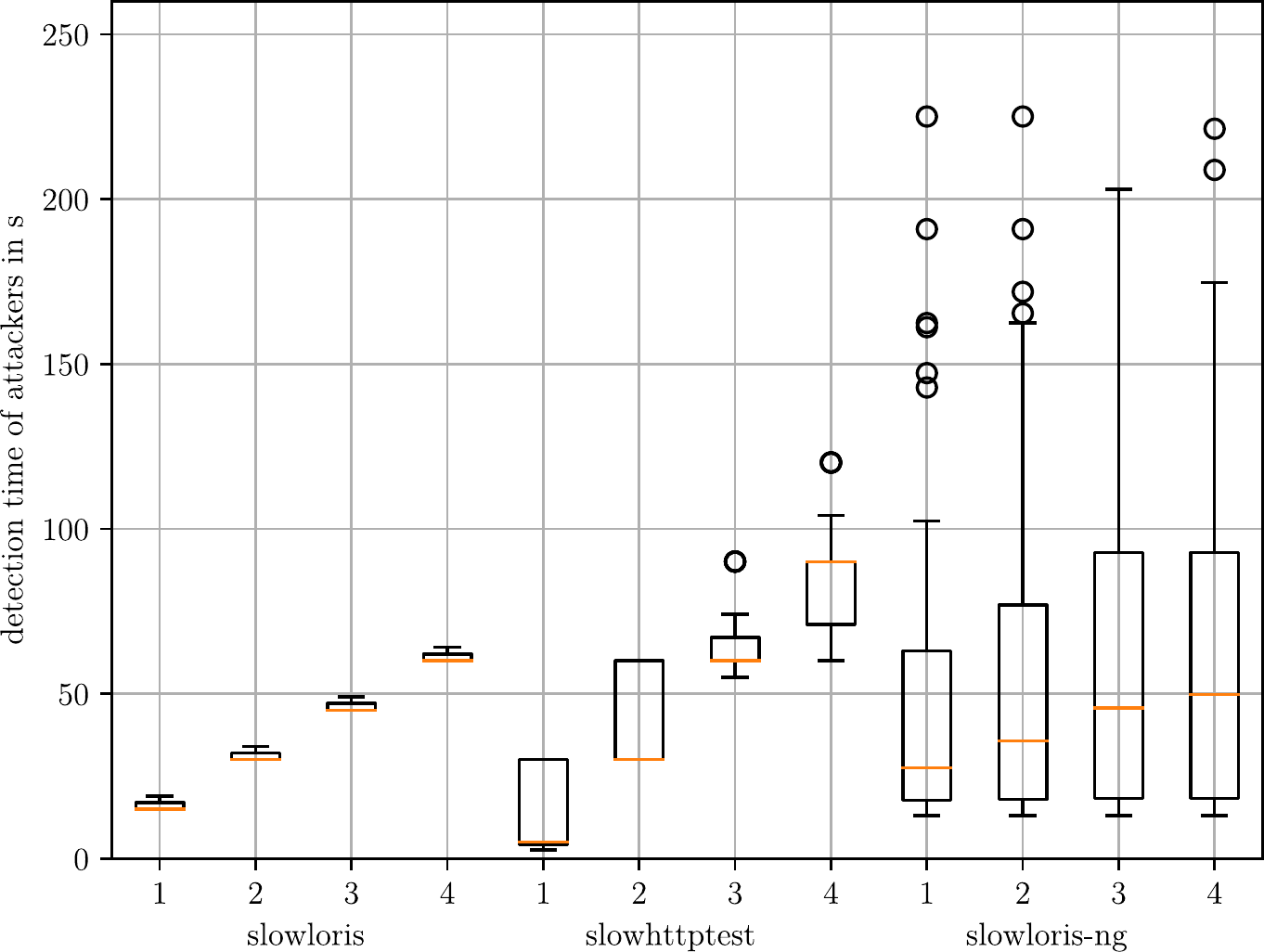}
  }
  \caption{Evaluation results for balanced accuracy (left) and detection times (right) for LPR without TCP handshake on SUEE8 dependent on strikes (p=0.77687Hz). More strikes result in a higher balanced accuracy but also much higher detection times.}
   \label{fig:strikes}
\end{figure}

For all these evaluations one suspicious packet is enough to deem a client an attacker. For the SUEE8 data set, this means that a client has to send only one suspicious packet in more than one week. The accuracy can be improved when several strikes, i.e. suspicious packets are necessary for an attacker classification. We conducted all tests with one, two, three, and four strikes. As an example, Figure~\ref{fig:strikes} shows our results for low packet rate without TCP handshake and a fixed threshold of 0.77687Hz for all measurements. The left figure shows, that the balanced accuracy can be improved when the amount of strikes necessary for a detection is increased. However, as can be seen in the right figure, this extends the detection time extensively, from a mean of 16 seconds to 61 seconds for slowloris, from 12 seconds to 86 seconds for slowhttptest, and from 53 to 67 seconds for slowloris-ng, this means a 281\%, 616\% and 26\% increase in detection time. Therefore, the comparably low increase in accuracy does result in impractical detection times.


\subsection{Discussion}

In line with our expectations, the schemes provide better results for the SUEE8 data set compared to the results of the ICSI data set. This is because the identification thresholds were determined using training data more similar to the former load.
This might indicate the importance of taking the actual deployment network into account, further tests need to be conducted to come to a definitive conclusion.
The evaluation shows varying results for the different schemes. In general, LC, MPR, and PRV are only able to detect slowhttptest. The TCP handshakes have to be taken into account when using these schemes. LC is both more accurate and faster than the other two. LPR, PDU, and the combined scheme LPR-PDU show much more potential. The best results in our measurements are achieved with LPR-PDU while also including TCP handshakes. We reached a balanced accuracy of 0.992 when using this scheme.
The varying threshold values per attack introduce additional complexity for successful defenses. 
We recommend less aggressive thresholds that detect simpler attacks as a default, and then dynamically ramp up to more aggressive thresholds in case the initial mitigation is not successful.


\section{Conclusion}
\label{sec:conclusion}

Slow HTTP attacks differ quite a bit from flooding attacks and their identification and mitigation can lead to a high management effort of the network infrastructure. We developed several concepts based on light-weight flow-based analysis of network traffic that can identify attackers and help to exclude them from the network.
Our analyses showed that a network-based defense approach against slow attacks is actually feasible and should be considered as part of a defense strategy for network providers. The accuracy of the schemes is not high enough to leave the system active all the time but it is very effective as part of a reactive defense system once ongoing attacks have been identified.
The attack tools we used are only able to conduct attacks based on the HTTP protocol. However, as we did not use any scheme that is dependent on application layer data, our mechanisms should also be able to protect other TCP-based application layer protocols that are vulnerable to slow attacks.

Except for the number of strikes, where we also considered the detection time, we choose the best scheme and threshold solely based on the balanced accuracy. However, the detection time should be considered as well. For future work, we would like to implement a system, that can dynamically alter thresholds based on the current network data to increase robustness and adaptiveness of the system.
With slow HTTP attacks, we are faced with the necessity to analyze large amounts of online traffic data efficiently. Detection time and accuracy of the mitigation system are in sharp contrast to each other and the right middle ground needs to be found. Therefore, for future work, we also plan to deepen our analysis concerning this compromise.

In this work, we also introduced slowloris-ng, a more sophisticated attack tool that is harder to detect by network operators due to its more randomized behavior. The effectiveness of attack tools for slow DDoS attacks relies on using a lot less resources than the system under attack. 
Slowloris-ng does require slightly more resources than slowloris for the same attack effectiveness (given no mitigation system is present) as the predictability of slowloris stems from its implementation optimized for highest impact. A formal analysis of the trade-off between attack effectiveness and detectability will be one of our next steps.

As part of our analysis, we compiled a new data set based on real-world traffic traces of one and eight days of benign traffic combined with attack traffic of three different attacks, marked in the data sets by their IP addresses. The data sets include traffic from nearly 10,000 unique IP addresses. We hope that by making these data sets publicly available without any restrictions, we can help others to both replicate our results and to enable the development of more efficient slow DDoS attack detectors.
\newpage

\section*{Acknowledgment}

We like to thank the Student Union of Electrical Engineering (Fachbereichsvertretung Elektrotechnik) at Ulm University and Philipp Hinz in particular for providing the necessary data.
This work was supported in the bwNET100G+ project by the Ministry of Science, Research and the Arts Baden-W\"urttemberg (MWK). The authors alone are responsible for the content of this paper.

%
%
\bibliographystyle{abbrv}
\bibliography{references}
%
\end{document}